\documentclass[twocolumn,amsmath,amssymb,floatfix,prl,superscriptaddress,showpacs]{revtex4}
\usepackage{amsmath,amssymb,natbib,bm,graphicx,url,epsfig}
\usepackage[ansinew]{inputenc}

\newcommand{\be}{\begin{equation}}
\newcommand{\ee}{\end{equation}}
\newcommand{\bes}{\begin{equation}\begin{split}}
\newcommand{\ees}{\end{split}\end{equation}}

\newcommand{\vc}[1]{\mathbf{#1}}

\newcommand{\abs}[1]{\left|#1\right|}

\begin{document}
\title{Pair tunneling through single molecules}
\author{Jens Koch}
\affiliation{Institut f\"ur Theoretische Physik, Freie Universit\"at Berlin, Arnimallee
14, 14195 Berlin, Germany}
\author{M.E.\ Raikh}
\affiliation{Department of Physics, University of Utah, Salt Lake City, 
UT 84112, USA}
\author{Felix von Oppen}
\affiliation{Institut f\"ur Theoretische Physik, Freie Universit\"at Berlin, Arnimallee
14, 14195 Berlin, Germany}
\date{October 10, 2005}
\begin{abstract}
By a polaronic energy shift, the effective charging energy of molecules 
can become negative, favoring ground states with even numbers of 
electrons. Here, we show that charge transport through such molecules 
near ground-state degeneracies is dominated by tunneling of electron 
pairs which coexists with (featureless) single-electron cotunneling. 
Due to the restricted phase space for pair tunneling, the current-voltage 
characteristics exhibits striking differences from the conventional 
Coulomb blockade. In asymmetric junctions, pair tunneling can be used for 
gate-controlled current rectification and switching.
\end{abstract}
\pacs{73.63.-b,72.10.Bg,81.07.Nb}

\maketitle

\emph{Introduction.}---Electronic transport through single molecules
is distinguished from transport through quantum dots by the coupling to
well-defined vibrational degrees of freedom. This coupling has two principal
consequences: First, it leads to the emergence of vibrational sidebands in
the current-voltage ($IV$) characteristics.  This phenomenon occurs at large
voltages which exceed the vibrational frequency, $\omega$. Its theoretical
description was pioneered in Refs.~\cite{glazman2,wingreen} and extended
in, e.g., Refs.~\cite{flensb1,aleiner,koch3}, following experiments \cite{park,ho,yu,pasupathy} on transport
through molecular junctions.

Second, the coupling to molecular vibrations induces a polaron shift 
and can lead  to a {\it negative} effective charging energy $U$. In physics, 
the concept of negative-$U$ centers was first pointed out more than
three
decades ago \cite{Anderson}, and is realized in many amorphous semiconductors. 
In chemistry, the scenario of negative $U$ is known as ``potential inversion" \cite{evans}.
An important ingredient in realizing this scenario in molecular junctions may be a
reduction of the true molecular charging energy by screening due to metallic
electrodes \cite{kubatkin} or an electrolytic environment \cite{schoenenberger}. Unlike vibrational sidebands,
negative $U$
manifests itself in transport through molecules already at low voltages. The
prime
manifestation of negative $U$ studied to date \cite{alexandrov1,alexandrov2,cornaglia1,arrachea,mravlje} concerns the Kondo
transport
at very low temperatures.

In general, finite on-site interaction (of any sign) opens a new transport
channel between molecule and leads. Namely, at finite $U$, {\em two} electrons can hop
onto the molecule {\em simultaneously}. But it is only for negative $U$ that this
process can dominate the transport. Indeed, negative $U$ favors even electron occupation
numbers of the molecule. A Kondo resonance occurs when two states, whose 
occupation numbers differ by two, are degenerate. Then, a virtual pair transition assumes 
the role of the spin-flip \cite{coleman}, leading to the formation of Kondo correlations 
in the ground state. However, the fact that the relevant pseudospin degree of freedom 
is associated with charge makes this Kondo state rather fragile. The underlying reason for 
this was elucidated by Haldane \cite{Haldane}, who pointed out that for the charge Kondo 
effect, any deviation from degeneracy acts as a Zeeman field on the pseudospin and, thus, 
suppresses the Kondo correlations. This observation has a drastic consequence for transport. 
For the conventional (spin) Kondo effect, the low-temperature conductance eventually reaches the
unitary value over a wide range of gate voltages. By contrast, for the charge Kondo effect the unitary
value is only achieved {\em precisely} at the resonance gate voltage. This fragility of the charge
Kondo effect in transport through molecules with negative $U$ was illustrated by several recent
numerical simulations \cite{cornaglia1,arrachea,mravlje}. The main message of the present paper is that
negative $U$ strongly
affects the transport through a molecule even in the rate-equation regime at high temperatures, 
where Kondo correlations are irrelevant. We show that in this regime, 
a negative $U$ leads to a unique scenario for the passage of
current through the molecule, which we study analytically at all gate voltages and biases.

\begin{figure}
	\centering
		\includegraphics[width=0.95\columnwidth]{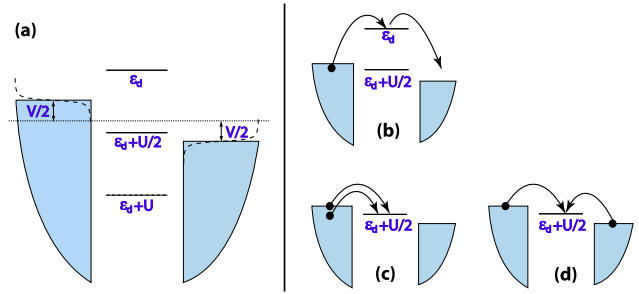}
	\caption{(a) Level configuration of the negative-$U$ model. Shown are the one-particle energies for the singly-occupied dot ($\varepsilon_d$), the doubly-occupied dot ($\varepsilon_d+U/2$), and the energy for which holes can propagate through the doubly-occupied dot ($\varepsilon_d+U$). The right panel illustrates the relevant types of processes: (b) cotunneling, (c) and (d) pair tunneling.\label{fig:levels}}
\end{figure}
\emph{Effective Hamiltonian.}---Consider a molecule with a single spin-degenerate single-particle
orbital $\varepsilon_d$ dominating the transport between two metallic electrodes. As $\varepsilon_d$ 
is sweeped by a gate voltage, the
electron occupation $n_d$ of the molecule switches from empty ($n_d=0$) to doubly
occupied ($n_d=2$) at
the resonant gate voltage defined by the condition $2\varepsilon_d+U=0$ 
\footnote{In the presence of additional orbitals, this remains true as long as 
$|U|$ is smaller than the level spacing.}. (We
measure all energies from
the Fermi energy.) Due to the negative $U$, single occupation ($n_d=1$) of the
molecule is unfavorable
at any gate voltage. A schematic of the corresponding configuration of
one-particle energies near
resonance is shown in Fig.~\ref{fig:levels}(a). Sequential tunneling of single
electrons is clearly
exponentially suppressed at bias voltages $|eV| \ll |U|$. Instead, the dominant
sequential transport
processes near resonance involve the coherent transfer of {\it electron
pairs} \footnote{We define sequential-tunneling processes as involving real occupation of the molecule as opposed to
cotunneling during
which the molecule is only virtually occupied.}. Representative pair-tunneling
processes are illustrated in Figs.~\ref{fig:levels}(c) and (d). It is important
to realize that in addition to processes in which the
two electrons enter the molecule from the same lead, the electron pair on the
molecule can be created
(or annihilated) by  two electrons tunneling in from (or out to) opposite leads.
In parallel with these
sequential pair-tunneling processes, single-particle cotunneling processes [see Fig.~\ref{fig:levels}(b)] also
take place.
We find that while both processes are of the same order, they can be easily
distinguished since
the single-particle cotunneling contribution does not exhibit any structure near
the resonance gate voltage.

To capture all eight pair-tunneling processes systematically, we perform a
Schrieffer-Wolff transformation \cite{schrieffer}.
We start from the Hamiltonian $H= H_\text{mol}
+H_{\rm vib}+H_\text{leads} + H_\text{i}$ \cite{glazman2,aleiner}, where the separate
contributions correspond to the electronic
molecular orbital $H_\text{mol}= \varepsilon_{d}n_d + U
n_{d\uparrow}n_{d\downarrow}$, to the molecular vibrations
$H_{\rm vib} = \hbar\omega b^\dag b + \lambda\hbar\omega(b^\dag+b)n_d$, to the
leads  $H_\text{leads}= \sum_{a=L,R}\sum_{\vc{p},\sigma}
\epsilon_\vc{p}c^\dag_{a\vc{p}\sigma}c_{a\vc{p}\sigma}$, and to the tunneling
between leads and molecule, $H_\text{i}= \sum_{a=L,R}\sum_{\vc{p},\, \sigma} \left( t_a
c^\dag_{a\vc{p}\sigma} d_\sigma + \text{h.c.}\right)$. Here, the operator
$d_\sigma$ annihilates an electron with spin $\sigma$ on the molecule,
$c_{a\vc{p}\sigma}$ annihilates an electron in lead $a$ ($a=L,R$) with momentum
$\vc{p}$ and spin $\sigma$, vibrational excitations are annihilated by $b$, and
$t_{L,R}$ denotes the tunneling matrix elements. The electron-phonon coupling (with coupling constant $\lambda$)
can be eliminated as usual by a canonical transformation
\cite{aleiner,glazman2}, which implies renormalizations of the tunneling
Hamiltonian $t_a\to t_ae^{-\lambda(b^\dag-b)}$, of the orbital energy
$\varepsilon_d \to \varepsilon_d-\lambda^2\hbar\omega$, and of the charging
energy  $U\to U-2\lambda^2\hbar\omega$. It is this last renormalization which
opens the possibility of a negative effective charging energy. 
The underlying reason for this renormalization is that the energetic polaron shift is 
proportional to the excess charge of the molecule {\it squared} which is of the same form as the 
Coulomb charging energy.
To be specific,
we
focus on temperatures and biases where only the vibrational ground state is
populated. In this case,
we obtain an effective Hamiltonian $H_{\rm eff}=H_{\rm mol} + H_{\rm leads} + H_\text{i}$
with negative effective charging energy
$U$ and Franck-Condon suppressed tunneling matrix elements $t_a\to t_a
e^{-\lambda^2/2}$ (absorbed into the
definition of $t_a$ in the following).

Performing the Schrieffer-Wolff transformation on $H_{\rm eff}$ in the usual way \cite{fedro},
we eliminate $H_\text{i}$ to lowest order
and obtain a transformed Hamiltonian
\begin{equation}
    H_\text{SW} = H_{\rm mol} + H_{\rm leads} + H_{\rm dir,ex} + H_{\rm pair},
\end{equation}
where we ignore a term which merely renormalizes $\varepsilon_d$. For positive $U$, one only retains 
\begin{align}
&H_{\rm dir,ex}=\frac{1}{2}\sum_{aa'\vc{p}\vc{p}'\sigma}\bigg[ \frac{t_at_{a'}^*}{\epsilon_{a\vc{p}}-\varepsilon_d}c_{a\vc{p}\sigma}^\dag c_{a'\vc{p}'\sigma}\\\nonumber
&+ t_at_{a'}^*\,M(\epsilon_{a\vc{p}})\left( d_{\bar{\sigma}}^\dag d_\sigma c_{a\vc{p}\sigma}^\dag c_{a'\vc{p}'\bar{\sigma}} - c_{a\vc{p}\sigma}^\dag c_{a'\vc{p}'\sigma}n_{d\bar{\sigma}}\right)+\text{h.c.}\bigg]
\end{align}
which describes the direct and exchange interactions between
molecule and lead. Here, we have introduced the abbreviations $M(\epsilon)= \frac{1}{\epsilon-\varepsilon_d}-\frac{1}{\epsilon-(\varepsilon_d+U)}$, and $\bar{\uparrow},\bar{\downarrow}=\downarrow,\uparrow$. With $U>0$ and in the absence of charge
fluctuations, transport predominantly proceeds via
single-electron cotunneling described by $H_\text{dir,ex}$. By contrast, for negative $U$ it is crucial to
retain the pair-tunneling terms
\be
H_{\rm pair} = \sum_{aa'\vc{p}\vc{p}'} t_at_{a'}\,M(\epsilon_{a\vc{p}})
d_{\uparrow}d_{\downarrow}c^\dagger_{a'\vc{p}'\downarrow}
   c^\dagger_{a\vc{p}\uparrow}+{\rm h.c.}
\ee
Obviously, $H_{\rm pair}$ contributes only for {\it nonzero} effective charging
energy $U$.

\emph{Rates.}---Using Fermi's Golden Rule, the pair Hamiltonian $H_{\rm pair}$
leads to the rate
\begin{align}\label{Wpairs}
&W_{0\to2}^{aa'} = \frac{\Gamma_a\Gamma_{a'}}{h}\int  d\epsilon\,M^2(\epsilon) f_a(\epsilon)f_{a'}(2\varepsilon_d+U-\epsilon),
\end{align}
for pairs of electrons tunneling onto the molecule. Here,
$\Gamma_a=2\pi\nu_a\abs{t_a}^2$ is the energy scale of single-particle
tunneling in junction $a$ for a constant density of states $\nu_a$. The
superscripts ($aa'$) denote
the leads from where the spin-up electron ($a$) and the spin-down electron
($a'$) originate. The analogous rates $W_{2\to0}^{aa'}$ for pairs
leaving the molecule are obtained from Eq.~\eqref{Wpairs} by replacing each
lead Fermi function $f_a$ with a factor $(1-f_a)$.

In the regime where single-particle occupation of the molecule is negligible, i.e.~$\abs{2\varepsilon_d+U}, \abs{eV}, k_BT \ll \varepsilon_d,\,\abs{\varepsilon_d+U}$,
the integral in Eq.~\eqref{Wpairs} approximately reduces to an integral over the Fermi functions alone, and can be expressed in terms of the function
 $F(\epsilon)=\epsilon/[\exp(\beta \epsilon)-1]$. For symmetric voltage splitting \footnote{The generalization to asymmetric voltage splitting is straightforward.}, i.e., $f_a(\epsilon)=f(\epsilon-eV_a)$ with $V_{L,R}=\pm V/2$, the explicit result for pair tunneling reads
\begin{align}\label{prates}
W_{\text{x}}^{aa'}&=\frac{\Gamma_a\Gamma_{a'}}{h}M^2(0)F\left[\pm\left( 2\varepsilon_d+ U+ eV_a+eV_{a'}\right)\right],
\end{align}
where the upper (lower) sign refers to $\text{x}=0\to2$ ($\text{x}=2\to0$).

The pair-tunneling rates, Eq.~\eqref{prates}, have several remarkable features.
(i) Pair tunneling with electrons originating from the same lead [$a=a'$,
Fig.~\ref{fig:levels}(c)] leads to bias-dependent rates. By contrast, pair
tunneling with electrons from different leads [$a\not=a'$,
Fig.~\ref{fig:levels}(d)] is bias independent, as the energy missing in the
lead with lower Fermi energy is exactly compensated for by the additional
energy available in the lead with higher Fermi energy. (ii) When pair tunneling
is energetically allowed, its rate is {\it proportional to the detuning} of the
pair state $2\varepsilon_d+U$ from the relevant Fermi energy. This is in sharp
contrast with rates for single-electron sequential tunneling, which are {\it
independent} of the energy of the single-particle state.
This unusual behavior of the pair-tunneling rates arises from the fact that only
the sum of the energies of the
two tunneling electrons is fixed by energy conservation, making the {\it phase
space} for pair tunneling proportional
to the detuning \cite{wilkinsbook}.
 We will see below that this leads to characteristic features of
pair tunneling both in the gate-voltage and bias dependence of the conductance.

Similarly, the rates for cotunneling from lead $a$ to $a'$ can be obtained from 
$H_{\rm dir,ex}$ as $W^{aa'}_{0\to0}=2 \Gamma_a\Gamma_{a'}F(eV_a)/h\varepsilon_d^2$,
including an explicit factor of 2 for spin. The corresponding rates $W^{aa'}_{2\to2}$ 
are obtained by the replacement $\varepsilon_d^2\to(\varepsilon_d+U)^2$. 

Having established the relevant processes and their rates, we now describe
transport outside the Kondo regime by the corresponding rate equations
\cite{beenakker3,averin3}. Since the occupation probability for the
singly-occupied molecule is negligible for $\abs{eV}\ll\abs{U}$, the stationary
rate equations reduce to $0=P_2W_{2\to0}-P_0W_{0\to2}$, with the
solution $P_0={W_{2\to0}}/[{W_{2\to0}+W_{0\to2}}]$, $P_2=1-P_0$.
Here, $W_{i\to f}$ denotes the total rate for transitions from initial state $i$
to final state $f$, i.e.,
$W_\text{x}=\sum_{a,a'} W^{aa'}_\text{x}$. The stationary current
$I=I^\text{pairs}+I^\text{cot}$,
evaluated in the left junction, involves contributions from pair tunneling and
cotunneling,
\begin{align}\label{ipairs}
I^\text{pairs}/e&=P_0 w_{0\to2}- P_2 w_{2\to0},\\
 I^\text{cot}/e &=P_0 v_{0\to0}+P_2 v_{2\to2}.\label{icot}
\end{align}
Here, the coefficients $v$ and $w$ are given by
\be\label{smallw}
v_{0\to0}=W^{LR}_{0\to0}-W^{RL}_{0\to0},\quad
w_{0\to2}=2W^{LL}_{0\to2}+W^{LR}_{0\to2}+W^{RL}_{0\to2}.
\ee
The factor of 2 in the last equation originates from the coherent transfer of
\emph{two} electrons in this pair-tunneling process. [All remaining
coefficients are obtained from Eq.~\eqref{smallw} by interchanging
``$2\leftrightarrow0$" in the subindices.]

\begin{figure}
	\centering
			\includegraphics[width=0.95\columnwidth]{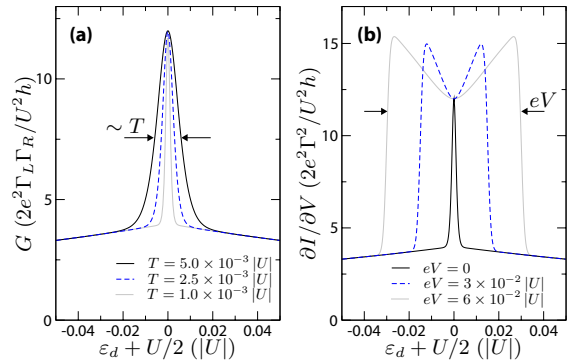}
	\caption{(color online) Conductance as a function of gate voltage, based on our analytical results.
	(a) Linear conductance $G$ for several temperatures.  The conductance curve exhibits a featureless background due to cotunneling and  a distinct peak of constant height and width $\sim T$ due to pair tunneling. (b) Conductance for several bias voltages at $T=1.0\times10^{-3}\abs{U}$ for a symmetric junction. With bias voltages $eV\gg k_BT$, the width of the double peak is given by $eV$.\label{fig:conduct}}
\end{figure}

\emph{Results.}---The rate equations provide a complete \emph{analytical} description
of the nonlinear current-voltage
characteristics through negative-$U$ molecules. Specifically, we find for the
linear conductance
\begin{align}
G=\frac{2e^2\Gamma_L\Gamma_R}{h}&\bigg[
\frac{U^2}{\varepsilon_d^2(\varepsilon_d+U)^2}\frac{\beta(2\varepsilon_d+U)}{2\sinh[\beta(2\varepsilon_d+U)]}\\\nonumber
+&f(-2\varepsilon_d-U)\frac{1}{\varepsilon_d^2} +
f(2\varepsilon_d+U)\frac{1}{(\varepsilon_d+U)^2} \bigg].
\end{align}
Here, the last two terms in brackets, which arise from cotunneling, give a slowly
varying background. By contrast,
pair tunneling, described by the first term in brackets, leads to a remarkable
conductance peak. As illustrated in
Fig.~\ref{fig:conduct}(a), its height $G_0=24e^2\Gamma_L\Gamma_R /U^2h$ exceeds twice the cotunneling background and is
temperature independent, while its width is proportional to $T$. This feature,
which is a direct consequence of the phase space for pair tunneling, should
provide
an unambiguous experimental signature of pair tunneling as opposed to ordinary
Coulomb blockade conductance peaks, whose integral strength is temperature
independent.

\begin{figure}
	\centering
  		\includegraphics[width=\columnwidth]{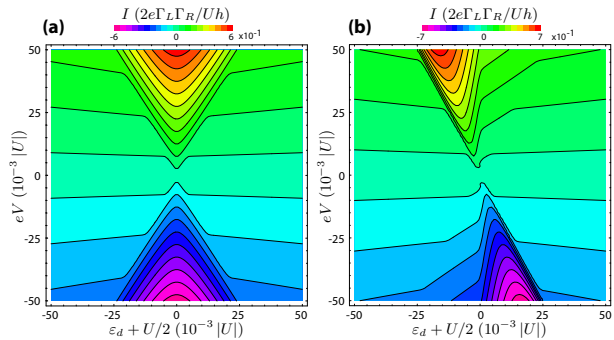}
	\caption{(color online) Current as a function of bias and gate voltage for (a) a symmetric junction with $\Gamma_L=\Gamma_R=k_BT$, (b) an asymmetric junction with $\Gamma_L=0.1k_BT$, $\Gamma_R=10k_BT$, and with $k_BT=0.5\cdot10^{-3}\abs{U}$. For asymmetric junctions, one observes the rectification effect due to pair tunneling. \label{fig:contours}}
\end{figure}

While the linear conductance is identical for symmetric and asymmetric devices,
their finite-bias behavior differs considerably. For large voltages
$\abs{eV}\gg k_BT$, we find for the pair tunneling current
\begin{align}\label{finiteV}
&I^\text{pairs} =\frac{e}{\hbar} \frac{16U^2}{(U^2-\delta^2)^2}\frac{2\Gamma_L\Gamma_R}{\Gamma_L^2\abs{\delta_+}+\Gamma_R^2\abs{\delta_-}+2\Gamma_L\Gamma_R
\abs{\delta}}\\\nonumber
&\times\bigg[ \Theta_+\Gamma_L\abs{\delta_+}(\Gamma_R\abs{\delta_-}+\Gamma_L\abs{\delta}) +
(L\leftrightarrow R,\, +\leftrightarrow -)\bigg],
\end{align}
where $\delta= 2\varepsilon_d+U$, $\delta_\pm=\delta\pm eV$, and $\Theta_\pm =
\theta(-\delta)\theta(\mp \delta_\pm)-\theta(\delta)
\theta(\pm \delta_\pm)$. For symmetric devices, Eq.\ (\ref{finiteV}) implies 
that the width of the conductance curve is
fixed by the bias voltage, as illustrated in Fig.~\ref{fig:conduct}(b). 
This behavior is in stark contrast to the conventional Coulomb blockade, where
one finds two symmetric sharp peaks at detunings $\pm V/2$ from resonance. This difference which also arises from the phase space for pair tunneling,
is further emphasized in Fig.~\ref{fig:contours}(a) for all bias and gate voltages.  

Devices with a significant asymmetry in the molecule-lead couplings exhibit a
striking asymmetry with respect to voltage inversion, see
Fig.~\ref{fig:contours}(b). We find in this case that pair tunneling causes 
strong current rectification, whose transmission direction can be
switched by a gate voltage.
To understand this rectification effect, consider an asymmetric device with
$\Gamma_R\gg\Gamma_L$, and suppose
that the Fermi level is higher in the left lead. Then, the pair-tunneling
current proceeds as follows: (i) A pair with electrons from opposite leads
jumps onto the molecule; (ii) the electron pair on the molecule is transferred
to the right lead. While (ii) is a fast process ($\sim\Gamma_R^2$), the current
is limited by process (i) with rate $\sim\Gamma_L\Gamma_R$. Then, the switching
by a gate voltage immediately follows from the rate for process (i), which is
exponentially suppressed only for $2\varepsilon_d+U>0$. By a similar analysis
for the opposite bias, one finds an exponential suppression of the pair current
for $2\varepsilon_d+U<0$, thus also explaining the rectification.
These features render molecules with negative $U$ interesting candidates for
devices with transistor-like characteristics.

\begin{figure}
	\centering
		\includegraphics[width=0.75\columnwidth]{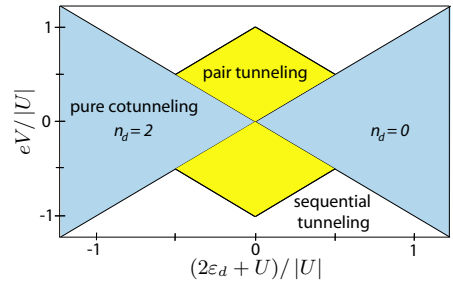}
	\caption{Stability diagram for the ground-state occupation $n_d$ and dominant transport modes as a function of gate and bias voltages for a symmetric device with negative $U$.\label{fig:phases}}
\end{figure}

\emph{Conclusions.}---Near the degeneracy point, transport through a
molecular junction with negative $U$ differs drastically from the
conventional Coulomb-blockade scenario at positive $U$. For negative $U$, current flow makes the molecule alternate in time between even occupation numbers, which is accomplished by pair tunneling. This is the dominant transport mode in the gate-voltage -- bias-voltage domain shown in Fig.~\ref{fig:phases}.

It is intriguing that the on-site attraction of two electrons makes 
pair tunneling through molecules
qualitatively similar to tunneling through a superconducting grain, considered in Ref.~\cite{hekking}.
However, in the case of molecules, the physical picture of
transport is more complex. This is because a pair can be created on
the molecule by electrons tunneling from {\em different} leads, as
illustrated in Fig.~\ref{fig:levels}(d). By contrast, for a grain 
with size larger than the superconducting correlation length, electron 
pairs enter and exit the grain only from the {\em same} lead, i.e.~only 
the processes in Fig.~\ref{fig:levels}(c) are responsible for the 
passage of current \footnote{Another difference between transport through negative-$U$ molecules and superconducting grains lies in the absence of BCS coherence factors in the former.}. As an important consequence of this difference, 
negative-$U$ molecules act as gate-controlled rectifiers. 

We thank Y.~Oreg and M. Wegewijs for useful discussions. 
This work was supported by 
NSF NER 057952 (MER), Sfb 658, and Studienstiftung d.\ dt.\ Volkes.

\end{document}